\documentclass[conference]{IEEEtran}
\IEEEoverridecommandlockouts
\usepackage{cite}
\usepackage{amsmath,amssymb,amsfonts}
\usepackage{textcomp}
\usepackage{xcolor}
\usepackage{graphicx}
\usepackage{booktabs}
\usepackage{tabularx}
\usepackage{algorithm}
\usepackage[noend]{algorithmic}
\usepackage{multirow}
\usepackage{color}
\usepackage{soul}
\usepackage{bbding}
\usepackage{pifont}
\usepackage{amsmath}
\usepackage{amsfonts}
\usepackage{subcaption}
\usepackage{arydshln}
\usepackage{url}
\usepackage{tikz}
\newcommand\copyrighttext{%
  \footnotesize \textcopyright 2024 IEEE. Personal use of this material is permitted.
  Permission from IEEE must be obtained for all other uses, in any current or future
  media, including reprinting/republishing this material for advertising or promotional
  purposes, creating new collective works, for resale or redistribution to servers or
  lists, or reuse of any copyrighted component of this work in other works.
  DOI: 10.1109/ICSC59802.2024.00022}
\newcommand\copyrightnotice{%
\begin{tikzpicture}[remember picture,overlay]
\node[anchor=south,yshift=10pt] at (current page.south) {\fbox{\parbox{\dimexpr\textwidth-\fboxsep-\fboxrule\relax}{\copyrighttext}}};
\end{tikzpicture}%
}
\newcommand{\oursystem}{$\mathsf{NORMY}$}
\newcommand{\cmark}{\ding{51}}
\newcommand{\xmark}{\ding{55}}

\newcolumntype{Y}{>{\centering\arraybackslash}X}

\def\BibTeX{{\rm B\kern-.05em{\sc i\kern-.025em b}\kern-.08em
    T\kern-.1667em\lower.7ex\hbox{E}\kern-.125emX}}
\begin{document}

\title{NORMY: Non-Uniform History Modeling for Open Retrieval Conversational Question Answering}

\author{\IEEEauthorblockN{Muhammad Shihab Rashid}
\IEEEauthorblockA{\textit{Computer Science and Engineering} \\
\textit{University of California Riverside}\\
Riverside, CA, USA \\
mrash013@ucr.edu}
\and
\IEEEauthorblockN{Jannat Ara Meem}
\IEEEauthorblockA{\textit{Computer Science and Engineering} \\
\textit{University of California Riverside}\\
Riverside, CA, USA \\
jmeem001@ucr.edu}
\and
\IEEEauthorblockN{Vagelis Hristidis}
\IEEEauthorblockA{\textit{Computer Science and Engineering} \\
\textit{University of California Riverside}\\
Riverside, CA, USA \\
vagelis@cs.ucr.edu}
}
\maketitle
\copyrightnotice
\begin{abstract}
\label{sec:abstract}
Open Retrieval Conversational Question Answering (OrConvQA) answers a question given a conversation as context and a document collection. A typical OrConvQA pipeline consists of three modules: a \textit{Retriever} to retrieve relevant documents from the collection, a \textit{Reranker} to rerank them given the question and the context, and a \textit{Reader} to extract an answer span. 
The conversational turns can provide valuable context to answer the final query. 
State-of-the-art OrConvQA systems use the same history modeling for all three modules of the pipeline. We hypothesize this as suboptimal. Specifically, we argue that a broader context is needed in the first modules of the pipeline to not miss relevant documents, while a narrower context is needed in the last modules to identify the exact answer span.
We propose NORMY, the first unsupervised non-uniform history modeling pipeline which generates the best conversational history for each module. 
We further propose a novel Retriever for NORMY, which employs keyphrase extraction on the conversation history, and leverages passages retrieved in previous turns as  additional context.
We also created a new dataset for OrConvQA, by expanding the doc2dial dataset. 
We implemented various state-of-the-art history modeling techniques and comprehensively evaluated them separately for each module of the pipeline
on three datasets: OR-QUAC, our doc2dial extension, and ConvMix. Our extensive experiments show that NORMY outperforms the state-of-the-art in the individual modules and in the end-to-end system.

\end{abstract}

\section{Introduction}\label{sec:intro}

Conversational Question Answering (CoQA) has recently attracted a lot of attention due to the widespread adoption of voice assistant platforms such as Siri, Alexa, and Google Assistant, and the advances in deep learning \cite{reddy2019coqa,choi2018quac,zaib2022conversational}. 
Given a text passage and a conversation, the goal of CoQA is to extract the answer to the last question of the conversation from the passage.
CoQA is an extension to Question Answering (QA) where the input is just one question instead of a conversation~\cite{kwiatkowski2019natural,voorhees1999trec,rajpurkar2016squad}.
However, in practice users do not provide an input passage when performing QA or CoQA.  This led to the newer problems of Open Retrieval QA (ORQA)~\cite{bajaj2016ms,cohen2018wikipassageqa,dhingra2017quasar} and Open Retrieval Conversational Question Answering (OrConvQA)~\cite{qu2020open}, 
where the input is a whole document collection.

State-of-the-art works on OrConvQA (also for ORQA) employs a pipeline of three modules ~\cite{qu2020open, qu2021weakly, fang2022open}.
The first one is a \textit{Retriever}, which retrieves a set of relevant passages from the collection. Both term-based (TFIDF/BM25) and embedding-based approaches may be used by the Retriever. A \textit{Reranker} module then re-ranks the already retrieved documents to better match the question, and finally, a \textit{Reader} module extracts an answer span from the re-ranked documents. Recent advances in transformers have produced pre-trained models like BERT which are highly effective in reader tasks~\cite{devlin2018bert}.

A key challenge in CoQA and OrConvQA is that the final user question may have co-references or ellipses, that is, some terms may refer to terms in the past conversation, while other useful contexts may be missing from the question. Further, previous (historic) turns of the conversation may add valuable context to the question being asked. 
Clearly, some of the past turns may be more useful than others as context for the last question. Blindly adding all turns may lead to a noisy history model. Including all turns may also be infeasible for some models like BERT, which can only support 512 tokens as the query and passage. 

Previous CoQA works propose different approaches to model the conversational history: some append all history turns to the final query making it a one big query and use it to retrieve the answer~\cite{choi2018quac,reddy2019coqa}, or use a backtracking algorithm which selects/disregards a particular history turn using deep reinforcement learning~\cite{qiu2021reinforced}, or rewrite the final query using the context of the whole conversation~\cite{vakulenko2021question,mele2021adaptive,su2019improving}. 
Previous OrConvQA works either use the previous 6 turns~\cite{qu2020open, qu2021weakly} or all turns with predicted answers~\cite{fang2022open} as context. 
\begin{figure*}[t]
\centering
  \includegraphics[width=0.6\textwidth]{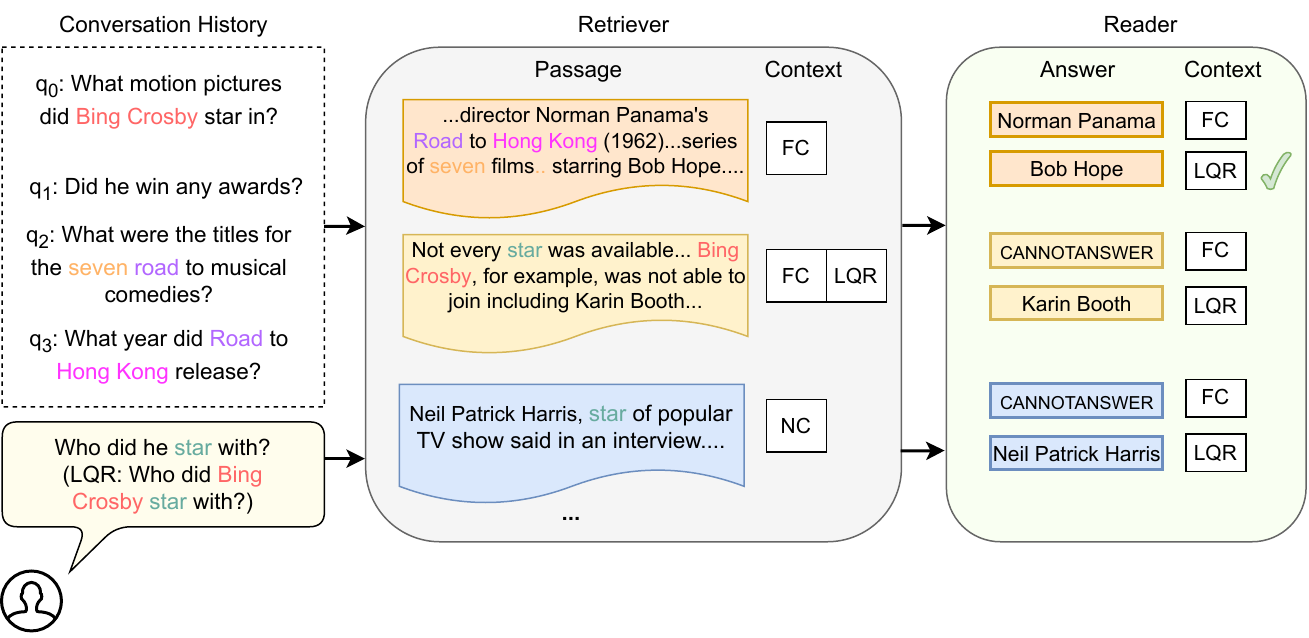}
  \caption[Intro]{Example of the impact of non-uniform conversational history modeling. Full Conversational Context (FC) retrieves the most relevant passages in the \textit{Retriever} module, while a narrower context, Last Question Rewrite (LQR) predicts the correct answer span in the \textit{Reader} module. 
  }
  \label{fig:intro_motiv}
\end{figure*}

Despite the different history modeling approaches of these previous works, they all use the same history model for all three modules of the pipeline. We hypothesize that this is suboptimal. Specifically, our hypothesis is that as we move towards the right of the \textit{Retriever}$\rightarrow$\textit{Reranker}$\rightarrow$\textit{Reader} pipeline and the number of the input passages (or documents) decreases, the history context should become shorter and more focused. That is, the Retriever should have access to broader context to not miss any relevant documents, whereas the Reader should have little context to help it identify the exact text span that  answers the user question. 

For example, in Figure~\ref{fig:intro_motiv} we see that modeling the history with Full Conversational Context (FC) returns the most relevant passage (top one) which has all the necessary information needed to answer the query.
In contrast, narrower context -- Last Question Rewrite (LQR) or No Context (NC) -- returns suboptimal passages that do not contain the answer. 
Once documents are retrieved, additional context (FC) may act as noise for the \textit{Reader} module, whereas more focused context (LQR) is able to extract the right span. 

In addition to using the same context, the state-of-the-art pipelines~\cite{qu2020open, qu2021weakly, fang2022open} and history modeling approaches~\cite{vakulenko2021question,mele2021adaptive,qiu2021reinforced} depend fully on training data to fine-tune the \textit{Retriever}, \textit{Reranker}, and \textit{Reader} modules. Finding quality training data for various domains of OrConvQA datasets is challenging. Ideally, a pipeline should be domain-agnostic and use appropriate history modeling to capture the context.
In this paper, we contribute in three ways towards solving the OrConvQA problems: (a) we propose {\oursystem}\footnote{\textbf{N}on-Unif\textbf{ORM} Histor\textbf{Y} Modeling}, the first unsupervised solution pipeline, (b) we build and publish a new dataset, and (c) we implement and experimentally compare various state-of-the-art history modeling algorithms for each of the three modules of the \textit{Retriever}$\rightarrow$\textit{Reranker}$\rightarrow$\textit{Reader} pipeline.

Our proposed system,  {\oursystem}, uses a \textit{non-uniform} history context for the three pipeline modules.
We also propose a novel history modeling algorithm for the \textit{Retriever} module that produces improved results over state-of-the-art baselines. Unlike previous approaches where the passages retrieved in previous turns are discarded, our Retriever algorithm considers past passages as candidates and proposes a ranking function that combines turn-based decay with context-based reranking of each passage. This ensures we do not miss an important passage due to noise being added in later turns.
Table~\ref{tab:introcomparison} summarizes the related work landscape, where we see that none of the previous work addresses all aspects of the problem. Only NORMY is question answering, open retrieval, conversational, performs history modeling, and it is non-uniform.

 \begin{table}[t]
  \caption{Comparison of selected tasks and datasets on the dimensions of Question Answering(QA), Open Retrieval(OR), Conversational (Conv), History Modeling (HM) and Non-uniform History Modeling (NHM)}
  \label{tab:introcomparison}
  \begin{tabular}{l|c|c|c|c|c}
    \hline
    Task/Dataset & QA & OR & Conv & HM & NHM\\
    \hline
    NQ~\cite{kwiatkowski2019natural}, SQuAD~\cite{rajpurkar2016squad} & \cmark & \xmark & \xmark & \xmark & \xmark\\
    TriviaQA~\cite{joshi2017triviaqa},\\MSMarco~\cite{bajaj2016ms},DrQA~\cite{chen2017reading} & \cmark & \cmark & \xmark & \xmark & \xmark\\
    CoQA~\cite{reddy2019coqa},\\QuAC~\cite{choi2018quac},ShARC~\cite{saeidi2018interpretation} & \cmark & \xmark & \cmark & \xmark & \xmark\\
    HAE~\cite{qu2019bert}, RL~\cite{qiu2021reinforced}, RW~\cite{vakulenko2021question} & \cmark & \xmark & \cmark & \cmark & \xmark\\
    OrConvQA~\cite{qu2020open}, d2d~\cite{feng2020doc2dial} & \cmark & \cmark & \cmark & \xmark & \xmark\\
    \textbf{\oursystem[ours]} & \cmark & \cmark & \cmark & \cmark & \cmark\\\\
  \hline
\end{tabular}
\end{table}
\begin{figure*}[t]
\centering
  \includegraphics[width=0.7\textwidth]{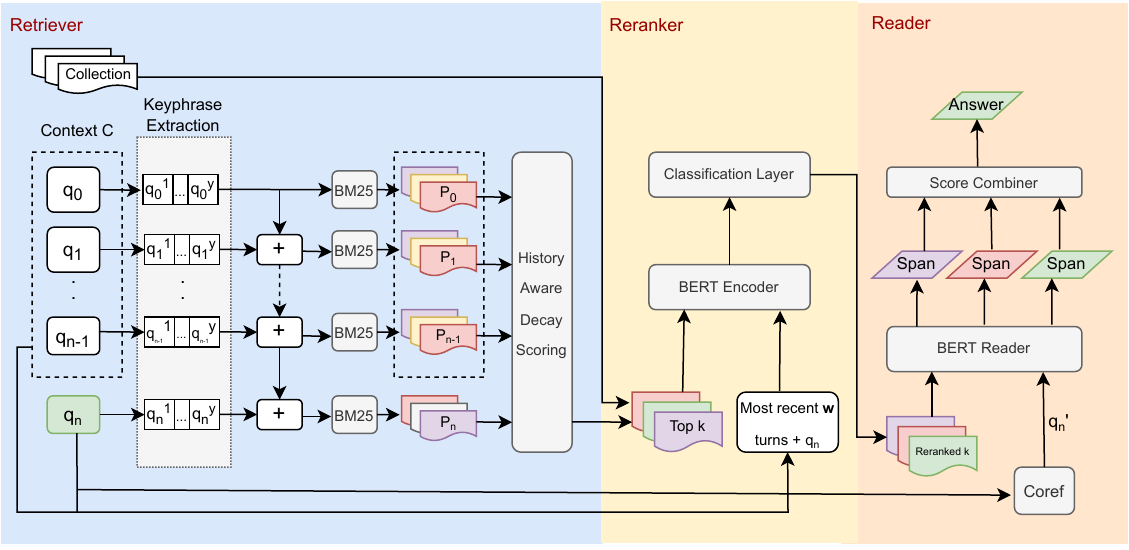}
  \caption[Intro]{The architecture of \oursystem. The input is the current question $q_n$, all history questions $q_i^{n-1}$, and the document collection $D$. The \textit{Retriever} module models the history using keyphrase extraction per history turn and retrieves passages $P_0 \cdots P_k$ using BM25. Our novel History Aware Decay Scoring module refines all returned passages and outputs top-k. The \textit{Reranker} reranks the passages using most recent $w$ turns and \textit{Reader} uses coreference resolution to rewrite the last query $q_n$ and outputs the best answer span combining all three modules' scores.} 
  \label{fig:overview}
\end{figure*}
 We evaluate our individual modules and the overall pipeline using 
 three varied datasets. First, we use the \textit{ORQUAC} dataset~\cite{qu2020open}, which is an extension of the CoQA dataset~\cite{reddy2019coqa}. A drawback of this dataset is that it does not portray natural dialogue conversation, as the chat is limited to asking questions and getting answers. Thus, we selected the \textit{doc2dial}~\cite{feng2020doc2dial} dataset for additional evaluation. However, this dataset is not created for the open retrieval conversational QA task as there are only a small number of documents as a corpus and the focus was to generate natural language answers and not text spans. For that, we created an updated \textit{doc2dial} dataset, which we call \textit{doc2dial-Or}. Third, we conduct experiments on ConvMix~\cite{christmann2022conversational}, where the corpus includes single-sentence passages and the history turns contain fewer co-references than previous datasets mentioned. NORMY outperforms the state-of-the-art in all three datasets.
 In summary, we make the following contributions in this paper: 
\begin{itemize}
    \item We identify the problem of uniform history modeling in conversational QA and  propose the first end-to-end pipeline for OrConvQA that uses non-uniform history modeling.
    \item We propose NORMY, a new unsupervised non-uniform universal history modeling pipeline. NORMY employs a novel history modeling approach for the Retriever module, which builds on keyphrase extraction principles, and leverages returned passages from previous history turns. 
    \item We perform an extensive comparison and analysis of various history modeling techniques for each module of the pipeline, on three diverse and structurally different datasets, and show that using the same modeling is suboptimal.
    \item We expand the \textit{doc2dial} dataset for the OrConvQA task and make our full source code and dataset available to the community~\footnote{\url{https://github.com/shihabrashid-ucr/normy}}.
\end{itemize}

The rest of the paper is organized as follows. An overview of the problem and solution are presented in Section~\ref{sec:our_system}. We present the details of NORMY including our novel Retriever algorithm in Section~\ref{sec:modules}. We present the datasets and results of our experimental evaluation in Section~\ref{sec:exp}. We discuss the related work in Section~\ref{sec:related_work} and finally conclude in Section~\ref{sec:conclusion}. 
\section{Problem Definition and Overview of NORMY}\label{sec:our_system}
\textit{Problem Definition.}
The input to the OrConvQA problem is a question $q_n$, the conversational history $C = q_0,\cdots,q_{n-1}$, and a document collection $D$. As in previous work, the history does not contain the answers to the questions, we also assume no access to the ground truth answers~\cite{qu2020open}. The output is an answer span $a_n$, extracted from one of the documents in $D$, which best answers $q_n$. The solution pipeline is shown in Figure~\ref{fig:overview}.
There are two key decisions we have to make for each module. First, pick what algorithm to employ (e.g. BM25 \cite{robertson1995okapi} or BERT~\cite{devlin2018bert} and so on) and with what parameters. Second, define what conversational context $C$ to input to the algorithm.  
In this work, we employ the state-of-the-art algorithm for each module and focus on the choice of conversational context for each module.

\textit{Overview of NORMY.}
NORMY, as shown in Figure~\ref{fig:overview}, generates a different model of the conversational history for each module of the pipeline.
Given the Retriever's history model discussed below and the collection, the \textit{Retriever} selects the top $k$ passages using BM25. 
Then, using a history of the last  \textit{w} turns, the \textit{Reranker} reranks the $k$ passages using transformer-based similarity measures. Finally, the transformer-based \textit{Reader} module models the history by rewriting the final query into a self-contained query, using coreference resolution, to find the best answer span. 
The answer span with the highest combined score from all three modules is the final answer.
 Note that our whole system is designed in an unsupervised fashion. There is no training data needed.

\section{Modules of NORMY}\label{sec:modules}

\subsection{Retriever}
The \textit{Retriever} module retrieves the \textit{k} most relevant passages from a document collection $D$, given a query $q_{n}$ and context $C$. 
We considered two types of search algorithms: classic Information Retrieval BM25-style ranking methods, and dense retriever methods. 
Although dense retriever approaches like ORQA~\cite{lee2019latent} and DPR~\cite{karpukhin2020dense} which use encodings of documents using ALBERT~\cite{lan2019albert}, have shown to provide better results, they require training data for fine-tuning. Their vanilla pre-trained models without training do not perform as well as BM25 (shown in Section~\ref{sec:exp}).
Further, BM25 is more scalable for large collections.
Hence, given that the main focus of this paper is history modeling, we picked BM25 for our Retriever.
Specifically, we index the documents using Lucene\footnote{https://lucene.apache.org/pylucene/}. Then we retrieve top \textit{k} documents using BM25, which is a term frequency based document ranking algorithm.

\textbf{History Modeling.} 
 NORMY's Retriever has two key novelties. First, we use a \textit{keyphrase extraction}-based candidate selection algorithm to identify the key context from the whole conversational history.
Second, we consider all passages returned by previous turns alongside passages returned by final turn as candidate passages, and rank them using \textit{history aware decay scoring} method to return top $k$.
\begin{algorithm}[t]
\caption{NORMYRetriever}
\begin{algorithmic}[1]
\label{alg:backe}
\REQUIRE Context $C, q_{n}$
\ENSURE SEL: Top \textit{k} returned passages
\STATE $SEL \leftarrow []$, $P \leftarrow []$
\FOR{\textit{each turn}  $i \in 1\cdots n$}
    \STATE $R(q_i) \leftarrow YAKE(q_{i})$
    \STATE $R(C) \leftarrow R(q_{0}) \cup \cdots \cup R(q_{i})$
    \STATE $P_i \leftarrow Retrieve_k(R(C))$~//top-k by BM25
        \STATE $P \leftarrow P \cup P_i$
    \FOR{\textit{each passage}  $p \in P_i$}
        \STATE $Compute~score~S_{rt}(p)~using~Eq.~2$
    \ENDFOR
\ENDFOR
\STATE $SEL \leftarrow$ SelectTopk$(P)$~//based on $S_{rt}(p)$
\RETURN $SEL$
\end{algorithmic}
\end{algorithm}

Our retrieval algorithm is shown in Algorithm~\ref{alg:backe}.
We extend the keyphrase extraction algorithm YAKE~\cite{campos2018text} to select $y$ best keywords per conversation turn  
using the YAKE formula shown in Equation~\eqref{eq:yake}. YAKE considers features like the casing of the word, word positions, word frequencies, word relatedness to context, etc. to assign a score $S(b)$ to each word $b$.
\begin{equation}
    \label{eq:yake}
    S(b) = \frac{W_{Rel} \cdot W_{Pos}}{W_{Case} + (W_{freq}/W_{Rel}) + (W_{DifS}/W_{Rel})}
\end{equation}
where $W_{Rel}$ is the relatedness to context score, $W_{Pos}$ is the word position score, $W_{Case}$ is the word casing score, $W_{freq}$ is the word frequency divided by the sum of mean term frequency and standard deviation $\sigma$, and $W_{DifS}$ is calculated based on how many times a particular word appears in other sentences. The detailed equations of all the terms can be found in~\cite{campos2018text}. We compute the union $R(C)$ of the reformulated questions $R(q_0) \cdots R(q_n)$ to retrieve the top $k$ passages (line 4-5). Each passage returned has a BM25 score assigned to it. 

\textbf{History-Aware Decay Scoring.} 
After retrieving $k$ passages for the current turn $n$, we refine their scores by considering their similarity to the retrieved passages. Further, we assign less weight to passages returned from previous turns. 
Specifically, we use a decay weight $\lambda$ to update the scores of older passages. The passages returned from previous turns may be relevant for subsequent modules but they do not share equal weight to passages returned from current turn $n$. Next, to assess the relevance of each passage $P_{nj}$, $\{j=1 \cdots k\}$ from turn $n$, we compute the average pairwise similarity with passages returned in the previous turn $P_{(n-1)i}$, $\{i=1 \cdots k\}$. This ensures that the passages returned has relevance to the whole conversation. Passages returned from irrelevant conversation turns will be scored less. To compute the similarity, we use SBERT~\cite{reimers-2019-sentence-bert} to produce embeddings of passages and perform cosine similarity. We update the score of $P_{nj}$ using this similarity. Finally, we rank all the passages using updated scores $S_{rt}$ and select top $k$. The retriever score of a passage $p$ is shown in Equation~\eqref{eq:hads}.
\begin{equation}
    \label{eq:hads}
    \small
    S_{rt}(q_n, C, p) = max(BM(R(C\cup q_n), p) - \lambda,0) \cdot {\sum_{i=1}^{k}}sim(p, P_{(n-1)i})/k
\end{equation}
where \textit{BM(.)} is the BM25 score of a passage and \textit{sim(.)} returns semantic similarity between two passages.
\subsection{Reranker}
The \textit{Reranker} module reranks the retrieved top $k$ passages using transformer-based encoders and a neural network to compute passage relevance score. The transformer based Reranker augments BM25 ensuring an extra layer of passage relevance. As $k<<$\textit{total size of collection}, using a transformer encoder is inexpensive.
A Reranker has been shown to improve the overall performance of the end-to-end system with little additional cost~\cite{qu2020open,htut2018training}. However, we show that using the same history modeling as the previous module or using the context from all history turns do not give the best results as now we have grounded documents as evidence. Our experimental results show that using a context with a history window size $w$ works best.
The input to the module is the final query $q_n$, the context $C$, and $k$ passages retrieved by the Retriever.
A reranking score $S_{rr}$ is assigned to every passage.

\textbf{Encoder.} Our Reranker module uses BERT to encode the input representation. We use the last $w$ history turns before $q_n$ and concatenate them together to model the history.
We then concatenate the retrieved passage $p_j$, $j=\{1 \cdots k\}$ to the appended history turns to create the final input sequence $ (q_n, C, p_j) =$ [CLS] $q_{n-w}$ [SEP] $\cdots$ [SEP]$q_{n-1}$ [SEP]$q_n$ [SEP]$p_j$. 
We use the contextualized vector representation of the input sequence ${\nu}_{[CLS]}$, and use it as input to a fully connected feed-forward layer that classifies the given passage as either \textit{relevant} or \textit{non-relevant} and outputs a classification score $S_{rr}$:
\begin{equation}
    {\nu}_{[CLS]} = {W}_{[CLS]} BERT{(q_n, C, p_j)}_{[CLS]}
\end{equation}
\begin{equation}
    S_{rr} = P(Rel=1|q_n, C, p_j) \overset{\triangle}{=} softmax({\nu}_{[CLS]})
\end{equation}
where ${\nu}_{[CLS]} \in \mathbb{R}^{T}$, T is the model embedding dimension, which is 768, and $W_{[CLS]}$ is a projection of the $[CLS]$ representation to obtain the sequence representation ${\nu}_{[CLS]}$. We compute the score for each passage in top $k$ independently and rerank them based on $S_{rr}$. 
\subsection{Reader}
The \textit{Reader} module inputs the final query $q_{n}$, the context $C$ and the reranked passages $\{p_1, p_2 .... p_k\}$ and outputs a span from one of the passages as the answer.

\textbf{History Modeling.} 
As the documents have already been narrowed down using the conversational context in previous modules, we show that using a history modeling with full contextual information produces worse results than a history model that uses less context. This happens due to: 1) The passages already hold the necessary contextual information from the history, 2) Previous history questions in the context misdirects the BERT Reader model into predicting incorrect answer spans.
The naive idea would be to use just the final query as input. However, the final query is prone to co-references and ellipses as users will not use self-contained utterances in a natural conversation. Thus, we use a co-reference resolution model to generate a resolved final query ${q_{n}}'$ using the previous context.
We adapt the huggingface neural co-reference model~\footnote{https://huggingface.co/coref/} which uses two neural networks to assign a score to each pair of mentions (or co-references) in the input and their antecedents~\cite{clark2016improving}. 
The history turns $q_0$ to $q_{n-1}$ are concatenated and used to rewrite $q_n$ into ${q_n}'$.

\textbf{Encoder.} Our Reader module uses similar BERT architecture as the previous module to encode the input. The input sequence "$[CLS]{q_n}'[SEP]p_j$" is used to generate a representation of all tokens in the input.
Two sets of parameters, a start vector $W_s$ and an end vector $W_e$ are used to compute the score for the $m$-th token.
\begin{equation}
   {\nu}_{[m]} = BERT{(({q_n}', p_j))}_{[m]}
\end{equation}
\begin{equation}
    S_{s}({q_n}', p_j, [m]) = W_{s}{\nu}_{[m]} \quad  S_{e}({q_n}', p_j, [m]) = W_{e}{\nu}_{[m]}
\end{equation}
where $S_s$ is the start score of a token and $S_e$ is the end score. The span Reader score $S_{rd}$ is computed as the maximum score of each token being either the start or end token. The start token must appear before the end token in the input.
\begin{equation}
\small
    S_{rd}({q_n}', p_j, s) = \underset{[m_s], [m_e] \in ({q_n}',p_j)}{max}S_{s}({q_n}', p_j, [m_s]) + S_{e}({q_n}', p_j, [m_e])
\end{equation}
where $s$ is the answer span with the start token $[m_s]$ and end token $[m_e]$. The answer spans are re-ranked using the combined score of all three modules and the top answer is given as a prediction.
\begin{equation}
\small
    S({q_n}, C, p_j, s) = S_{rt}(q_n, C, p_j) + S_{rr}(q_n, C, p_j) + S_{rd}({q_n}', p_j, s)
\end{equation}
\section{Experimental Evaluation}\label{sec:exp}
\begin{table}[t]
\centering
  \caption{Dataset Statistics}
  \label{tab:datastat}
  \begin{tabular}{c|c|c|c}
    \toprule
     & ORQUAC & doc2dial-OR & ConvMix\\
    \midrule
    \# Dialogues & 771 & 661 & 1679\\
    \# Questions & 5571 & 4253 & 2284\\
    \# Avg tokens/qstn & 6.7 & 10 & 6.39\\
    \# Avg tokens/ans & 12.2 & 21.6 & 2.17\\
    \# Avg questions/conv & 7.2 & 6.4 & 5.00\\
    \# Passages & 11M & 11.6M & 5.94M\\
  \bottomrule
\end{tabular}
\end{table}
\subsection{Datasets}

We use three datasets with different conversation structures. The first dataset: ORQUAC~\cite{qu2020open} is an aggregation of three existing datasets: QuAC, CANARD~\cite{elgohary2019can}, and Wikipedia corpus that serves as a knowledge source for open retrieval. The Wikipedia corpus is a collection of 11 million passages that are created from splitting Wikipedia articles into 384 tokens. 

The second dataset is \textit{doc2dial-OR}, which was created by us, as an extension of \textit{doc2dial}~\cite{feng2020doc2dial} dataset, which consists of natural information-seeking goal-oriented dialogues that are grounded in documents. doc2dial has more complex questions than ORQUAC, associated with multiple sections of a document. 
However, this dataset is only grounded to 480 long documents collected from different government websites, which is not ideal for an open retrieval task. 
\textit{doc2dial-OR} extends doc2dial by having a much larger set of passages consisting of (a) 11 million Wikipedia passages\footnote{https://dumps.wikimedia.org/enwiki/20191020}, and (b) the 480 documents of doc2dial split into 384-token chunks. 
The second difference between doc2dial-OR from doc2dial is that we convert free-text ground truth answers to exact text spans from the gold passage in the dataset, to make it suitable for a span prediction task, as is the case for ORQUAC. 

The third dataset is ConvMix~\cite{christmann2022conversational}, which contains documents from heterogeneous sources: Wikipedia info boxes, tables, and text snippets (passages). They use the Wikipedia dump from 2022-01-31. To adapt this dataset for our task, we selected the 5.94 million textual snippets as our collection and the question turns in a conversation where the answer can be extracted from these text snippets.
The dataset statistics are shown in Table~\ref{tab:datastat}.
\begin{table*}[t]
\caption{Retriever Results}
\label{tab:retr_result}
\begin{tabular}{l|l|l|l|l|l|l|l|l|l|l|l|l} 
\hline
\multirow{2}{*}{Setting} & \multicolumn{4}{c|}{ORQUAC}                                          & \multicolumn{4}{c|}{doc2dial-OR}         & \multicolumn{4}{c}{ConvMix}                                \\ 
\cline{2-13}
                         & MRR             & R@1             & R@5             & R@10            & MRR             & R@1             & R@5             & R@10   & MRR & R@1 & R@5 & R@10         \\ 
\hline
No History               & 0.0312                & 0.0177                & 0.0516                & 0.0649                & 0.2564          & 0.2003          & 0.3320          & 0.3907  & 0.1027 & 0.0696 & 0.1449 & 0.186        \\ 
\hline
First Last               & 0.1174                & 0.0739                & 0.1757                & 0.2249                & 0.4283          & 0.3378          & 0.5523          & 0.6381  & 0.1586 & 0.109 & 0.2263 & 0.281        \\ 
\hline
Full History             & 0.1361          & 0.0785          & 0.1748          & 0.2178          & 0.4289          & 0.3376          & 0.5584          & 0.6433    & 0.1605 & 0.1077 & 0.2364 & 0.2911      \\ 
\hline
Fixed window      & 0.1222          & 0.0827          & 0.1744          & 0.2189          & 0.4292          & 0.3378          & 0.5584          & 0.6433  & 0.1605 & 0.1077 & 0.2364 & 0.2911        \\ 
\hline
Backtracking             & 0.1160          & 0.0726          & 0.1742          & 0.2238          & 0.4226          & 0.3301          & 0.5494          & 0.6364  & 0.1696 & 0.1169 & 0.2429 & 0.3104        \\ 
\hline
Rewriting                & 0.0516          & 0.0307          & 0.0814          & 0.1080          & 0.2605          & 0.2031          & 0.3369          & 0.3985   & 0.12 & 0.0827 & 0.1654 & 0.2123       \\ 
\hline
\textbf{NORMYRetr[ours]}     & \textbf{0.1662} & \textbf{0.1147} & \textbf{0.2367} & \textbf{0.2891} & \textbf{0.4687} & \textbf{0.3809} & \textbf{0.5906} & \textbf{0.6780} & \textbf{0.1757} & \textbf{0.119} & \textbf{0.2513} & \textbf{0.3139} \\
\bottomrule
\end{tabular}
\end{table*}
\begin{figure*}[t]
    \centering
    \centering
    \begin{subfigure}[b]{0.3\textwidth}
      \centering
      \includegraphics[width=\linewidth]{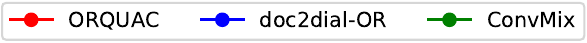}
    \end{subfigure}%
    
    \begin{subfigure}[b]{0.38\textwidth}
      \centering
      \includegraphics[width=0.75\linewidth]{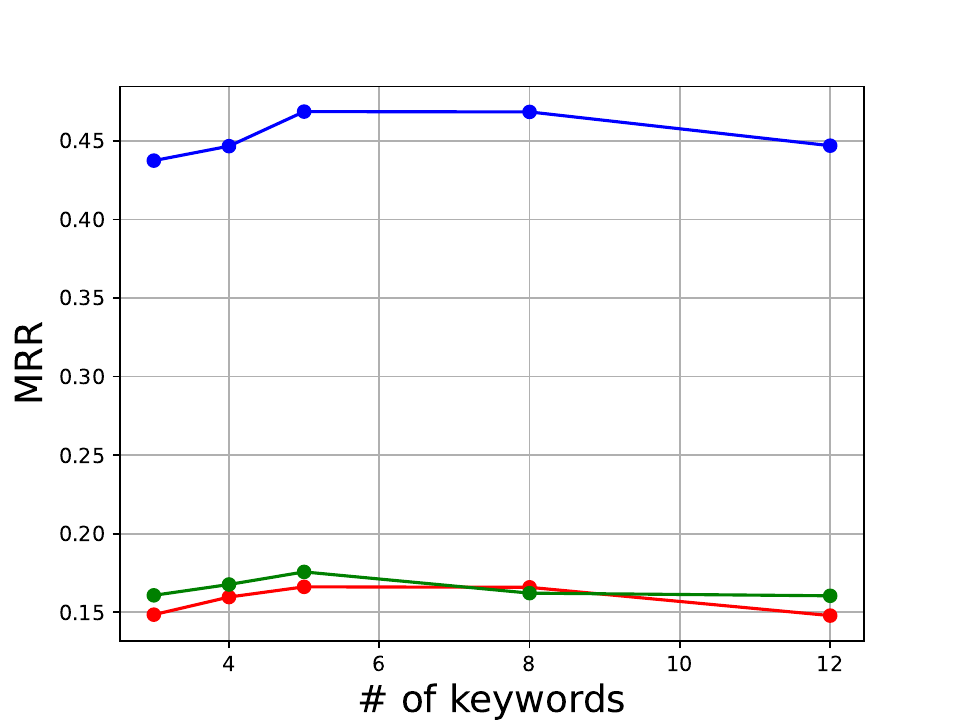}
      \caption{Retriever performance with $y$}
      \label{fig:retr_graph}
    \end{subfigure}%
    ~
    \begin{subfigure}[b]{0.38\textwidth}
      \centering
      \includegraphics[width=0.75\linewidth]{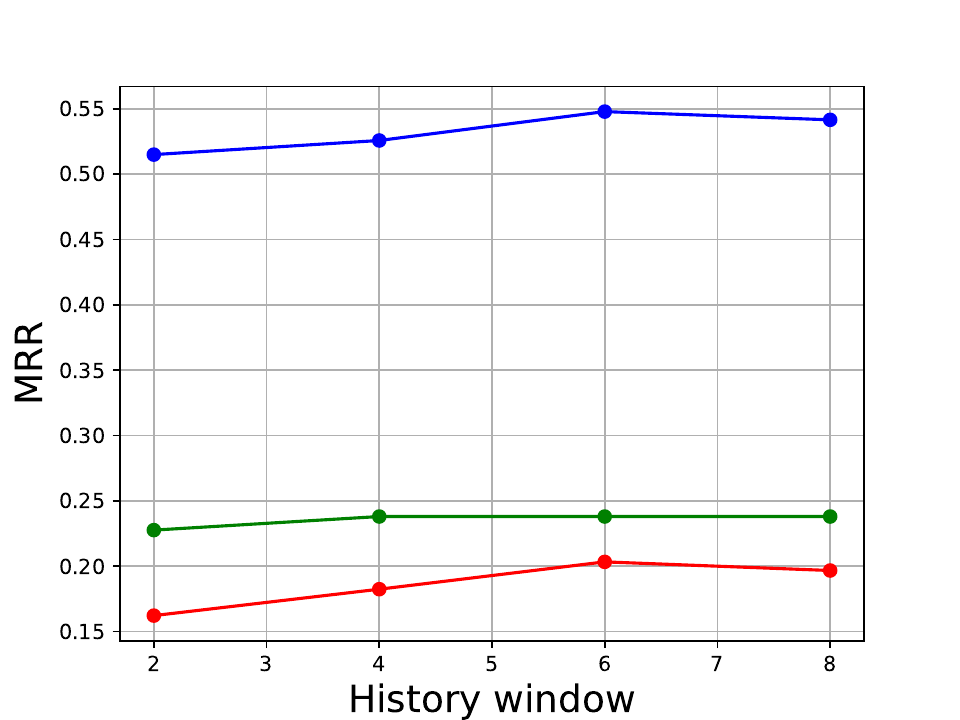}
      \caption{Reranker performance with $w$}
      \label{fig:rerank_graph}
    \end{subfigure}%
    \caption{(a) and (b) subgraphs show the impact of number of keywords $\textbf{y}$ and history window size $\textbf{w}$ for Retriever and Reranker.}
    \label{fig:graph_figs}
\end{figure*}
\subsection{Experimental Setup}

\textbf{Competing History Models.}
To the best of our knowledge, there are no fully unsupervised non-uniform history modeling approaches. There are supervised systems (OrConvQA~\cite{qu2020open}, WS-OrConvQA~\cite{qu2021weakly}) that uses history window of 6 for all the modules. ConvADR-QA~\cite{fang2022open} uses all history turns along with their predicted answers as context. However, they require annotated data to train the modules. We adapt their approaches to an unsupervised setting.
 There are also conversational closed retrieval systems (RL~\cite{qiu2021reinforced}, RW~\cite{vakulenko2021question}). We adapt such history modeling methods to an open-retrieval setting. Further, we also propose some standard history modeling techniques. Thus, we have identified the following baselines: \textbf{1) No History~\cite{choi2018quac}:} Where we do not perform any history modeling. The last question turn is used as input to each individual module. \textbf{2) First-Last:} We propose an intuitive history modeling baseline, where we define the history as the combination of the immediately previous user utterance and the first utterance of the conversation. 
\textbf{3) Full History:} With all previous turns concatenated. For the modules where we use transformer models with token limitation, we prune the earlier tokens if the total token size exceeds 384. The input sequence is $C = [CLS]q_0 [SEP] q_1$ $ [SEP] \cdots [SEP]q_n$ 
 \textbf{4) YAKE~\cite{campos2018text}:} Keyphrase extraction-based history modeling has not been used previously in any research work. We extract $y$ keyphrases per history turn and concatenate them with the last turn to create the input sequence.
 \textbf{5) Backtracking [RL]:} We adapt the \textit{immediate reward} based history selection proposed by Qiu et al.~\cite{qiu2021reinforced} for closed retrieval systems. We select a history turn if the similarity with previously selected history turns is greater than 0.5. For each history turn($i=1...n$) we calculate the immediate reward with SBERT sentence encoder.
 \textbf{6) Question Rewriting [RW]:} We adapt the algorithm of Vakulenko et al.~\cite{vakulenko2021question} for closed retrieval systems, which uses a question rewriting model to resolve ambiguous questions (co-references) into self-contained questions. Their model requires training data thus we use \textit{neuralcoref} to resolve the co-references of the final query using previous history turns. There are other query rewriting models like QReCC~\cite{anantha2020open} which also require annotated data.
 \textbf{7) Fixed Window [OrConvQA, WS-OrConvQA]~\cite{qu2020open, qu2021weakly}:} WS-OrConvQA is an improvement over OrConvQA but uses training data to learn weak supervision signals. Both models use a history window size $w$.  Window size 6 is shown to produce the best results for both. To select the baseline, we also compared different window sizes (2,4,6,8) in a small validation set and $w=6$ has given the best results. 
 \textbf{8) Fixed Window with Ans. [ConvADR-QA]~\cite{fang2022open}:} This model predicts the answers for each historical question with a teacher model using annotated query rewrites and appends the predicted answer to the context along with historical questions. For a fully unsupervised pipeline, we adapt this approach to predict an answer for each question using OrConvQA and append the answer to the context. The input sequence is $C = [CLS]q_0 a_0 [SEP] q_1 a_1$ $ [SEP] \cdots [SEP]q_n a_n$, where $a_n$ is the predicted answer for turn $n$.
\begin{table*}[t]
\caption{Reranker Results}
\label{tab:rerank_result}
\begin{tabular}{l|l|l|l|l|l|l|l|l|l|l|l|l} 
\hline
\multirow{2}{*}{Setting} & \multicolumn{4}{c|}{ORQUAC}                                          & \multicolumn{4}{c|}{doc2dial-OR}   & \multicolumn{4}{c}{ConvMix}                                       \\ 
\cline{2-13}
                         & MRR             & R@1             & R@5             & R@10            & MRR             & R@1             & R@5             & R@10   & MRR & R@1 & R@5 & R@10          \\ 
\hline
No History               & 0.1408                & 0.1033                & 0.1927                & 0.2891                & 0.4572          & 0.3592          & 0.5901          & 0.6780 & 0.2264 & 0.1834 & 0.285 & 0.3139          \\ 
\hline
First Last               & 0.1779                & 0.1491                & 0.2179                & 0.2891                & 0.5247          & 0.4514          & 0.6301          & 0.6780 & 0.2396 & 0.2052 & 0.2955 & 0.3139          \\ 
\hline
Full History             & 0.1996          & 0.1702          & 0.2402          & 0.2891          & 0.5401          & 0.4644          & 0.6415          & 0.6780   & 0.2405 & 0.2020 & 0.2944 & 0.3139        \\ 
\hline
\textbf{NORMY(Fixed $w$)}    & \textbf{0.2033} & \textbf{0.1767} & \textbf{0.2411} & \textbf{0.2891} & \textbf{0.5478} & \textbf{0.4747} & \textbf{0.6515} & \textbf{0.6780} & 0.2405 & 0.2020 & 0.2944 & 0.3139 \\ 
\hline
Backtracking             & 0.1954          & 0.1661          & 0.2361          & 0.2891          & 0.5325          & 0.4635          & 0.6368          & 0.6780   & 0.2403 & 0.2017 & 0.2940 & 0.3139        \\ 
\hline
Rewriting                & 0.1696          & 0.1344          & 0.2181          & 0.2891          & 0.4583          & 0.3604          & 0.5911          & 0.6780  & 0.2295 & 0.1873 & 0.2867 & 0.3139         \\ 
\hline
YAKE                     & 0.2008          & 0.1731          & 0.2387          & 0.2891          & 0.5396          & 0.4624          & 0.6377          & 0.6780  & \textbf{0.2418} & \textbf{0.2031} & \textbf{0.2946} & \textbf{0.3139}         \\
\bottomrule
\end{tabular}
\end{table*}
\textbf{Implementation Details.} NORMY is fully unsupervised without requiring any training data. The pre-trained models are implemented with the open-source library Huggingface. We index our document collection using PyLucene with \textit{StandardAnalyzer} as tokenizer
Indexing is done with term frequencies, document frequencies, and positions. Keyphrase extraction is done with open source library \textit{pke}~\cite{boudin:2016:COLINGDEMO}. For computing similarity, we use SBERT's pre-trained \textit{roberta-large} model. For \textit{Reranker} module, we use pre-trained BERT model finetuned on MSMARCO dataset~\cite{nguyen2016ms}. For our \textit{Reader} module we use a pre-trained \textit{bert-large} model finetuned on SQUAD dataset. We make our full code available to the research community.
\subsection{Retriever Results}
\textbf{Evaluation Methodology.} We use the commonly used \textit{Mean Reciprocal Rank (MRR)} and \textit{Recall (R@k)} methods to measure our retrieval performance. \textit{MRR} calculates how far down the ranking the first relevant document is on average, where higher is better. \textit{R@k} measures the fraction of times the correct document is found in the top $k$ predictions, where higher is better.\\
\textbf{Results.} We first experiment with different values of the number of keywords $y$ in the keyphrase extraction shown in Figure~\ref{fig:retr_graph} and find that $y=5$ performs best. We use $y=5$, $k=10$ and $\lambda=0.1$ in Table~\ref{tab:retr_result}. We found that using a larger $k$ increases the execution time of the system with a very small accuracy benefit. Figure~\ref{fig:graph_figs} shows our experimental results for different parameters used in NORMY.

We compare different history modeling techniques using BM25 in Table~\ref{tab:retr_result}. For all three datasets, we see that our NORMY Retriever performs significantly better than other baselines. This is due to a couple of factors: 1) We remove irrelevant information from each history turn rather than eliminating the history turn altogether, 2) We consider previously retrieved passages from previous history turns as candidate passages. We can also see that history models that use fewer contexts like Question Rewriting, First-Last, and No History perform significantly worse indicating we need more context while retrieving from millions of passages. 
\subsection{Reranker Results}
\textbf{Evaluation Methodology.} 
We use the same metrics as the Retriever as both of these modules produce top $k$ passages.\\
\textbf{Results.} From Table~\ref{tab:rerank_result} we see that the Reranker module significantly improves the ranks of relevant passages. Using a fixed window, which sits between a broader context like Full History and a narrower context like Rewriting produces the overall best results. We conduct experiments with different history window sizes $w$ and show the results in Figure~\ref{fig:rerank_graph}. Fixed window of 6 works best for two of our datasets, which is supported by the literature~\cite{qu2020open}. 
For ConvMix, we see that YAKE performs slightly better than all other methods and the results vary very little. This is because the passages in the collection are single sentences only, leading to multiple passages being relevant to the question. The \textit{Reranker} ranks such passages similarly whereas only one contains the gold answer. In the real world, it is unusual for the documents to be single sentences. Note that our hypothesis that the \textit{Reranker} needs less context than the \textit{Retriever} still holds, as  
YAKE has less context than Full History.
\begin{table}[t]
  \caption{Reader F1}
  \label{tab:reader_result}
  \begin{tabular}{c|c|c|c}
    \hline
     Setting & ORQUAC & doc2dial-OR & ConvMix\\
     \hline
    No History & 0.1557 & 0.1898 & 0.6785\\
    First Last & 0.0996 & 0.1291 & 0.4842\\
    Full History & 0.0845 & 0.1196 & 0.3711\\
    Fixed Window & 0.0848 & 0.1200 & 0.4859\\
    Backtracking & 0.1118 & 0.1541 & 0.5591\\
    \textbf{NORMY(Rewriting)} & \textbf{0.1774} & \textbf{0.2220} & \textbf{0.7393}\\
    YAKE & 0.1277 & 0.1551 & 0.67\\
  \bottomrule
\end{tabular}
\end{table}
\begin{table}[t]
  \caption{Entire Pipeline F1. $\ddagger$ means statistically significant improvement over baseline with $p < 0.5$.}
  \label{tab:framework_result}
  \begin{tabular}{c|c|c|c}
    \hline
     Setting & ORQUAC & doc2dial-OR & ConvMix\\
     \hline
    \textbf{NORMY[ours]} & $\boldsymbol{0.0782^\ddagger}$ & $\boldsymbol{0.1625^\ddagger}$ & $\boldsymbol{0.1723^\ddagger}$\\
    \textit{NORMY w/o decay} & $0.0668^\ddagger$ & $0.1323^\ddagger$ & 0.1490\\
    \textit{NORMY w/o sim} & $0.0695^\ddagger$ & $0.1431^\ddagger$ & 0.$1562^\ddagger$\\
    \hline
    OrConvQA~\cite{qu2020open},\\WS-OrConvQA~\cite{qu2021weakly}(BM25) & 0.0478 & 0.0955 & 0.1314\\
    OrConvQA,\\WS-OrConvQA (DPR) &  0.0466& 0.0948 & 0.1298\\
    ConvADR-QA~\cite{fang2022open} & 0.0454 & 0.0897 & 0.1244 \\
  \bottomrule
\end{tabular}
\end{table} 
\subsection{Reader Results}
\textbf{Evaluation Methodology.} We treat the evaluation of Reader module as a span selection task and adopt token level F1 as the evaluation metric. F1 calculates the similarity between the ground answer and the predicted span, where higher means better.\\
\textbf{Results.} From Table~\ref{tab:reader_result} we can see significant performance drops when more contexts are added for all three datasets. 
As the candidate passages have been reduced to 1, transformer-based reader model performs significantly better when only one query is used.
Narrower contexts like adding no history and question rewriting perform much better than broader context models further proving our hypothesis. Among them, question rewriting produces a better result.
For ConvMix we can see very high F1 scores for appropriate history models as the answers are on average two tokens only, which makes the \textit{Reader} model easily predict answer spans. However, history models with broader context have poor F1 scores for this same dataset indicating the model needs proper history models even in simpler scenarios.
\subsection{End-to-end Evaluation}
In this section, we compare our pipeline with the state-of-the-art models~\cite{qu2020open, qu2021weakly, fang2022open}, which either use a fixed window of 6 or full history with predicted answers uniformly in all modules. We see in Table~\ref{tab:framework_result} that SOTA models perform poorly in a fully unsupervised setting. We also see that, ConvADR-QA performs worse than OrConvQA, as in an unsupervised setting, a wrongly predicted answer could misdirect the context to retrieve irrelevant passages.
Our pipeline with non-uniform history modeling performs significantly better. SOTA models use fine-tuned dense retriever model (DPR) for their retriever module instead of BM25 which is used by NORMY. We use the vanilla pre-trained version of DPR here as we don't have access to training data. 
We also compare to a variant of OrConvQA that uses BM25 instead of DPR for completeness. We see that BM25 performs better than dense retriever models. Note that, uniformly using other baseline history modeling techniques evaluated in previous subsections does not produce better results than NORMY in the end-to-end pipeline. We do not show these in Table~\ref{tab:framework_result} for brevity.
\subsection{Ablation Studies}
The effectiveness of our model relies on some of the design choices we made. We investigate such choices our novel retriever \textit{NORMYRetr} has from equation~\eqref{eq:hads}. We present the ablation results in Table~\ref{tab:framework_result}. Specifically, we show two ablation settings as follows:

\textbf{NORMY w/o decay.} We showed that if we disregard previously returned passages we may miss out on some relevant information required for subsequent modules. However, if we gave the same weight to previous passages as passages returned from current turn \textit{n}, we see a degradation in performance. This is due to the current turn holding the most amount of information. Thus passages returned from the current turn should be given the most weight.

\textbf{NORMY w/o sim.} If a history turn is related to its previous turn, the passages returned will also have some similarities. Here, we disregarded the average pairwise similarity with the previous turn's returned passages from equation~\eqref{eq:hads}. We again see a decrease in model performance. By refining retriever scores with similarity score, we compute the relevance of passages with relevant conversational history.
The ablation studies further verify that both decay and similarity scores are crucial for NORMY to perform best.
\section{Related Work}
\label{sec:related_work}
\textbf{Machine Reading Comprehension (MRC).} MRC task typically includes a single-turn query where the answer grounds in a short passage. It started with TREC~\cite{voorhees1999trec} in the early days where the goal was to retrieve the appropriate passage for 200 factoid questions and advanced to recent high-quality datasets like NQ~\cite{kwiatkowski2019natural}, SQuAD~\cite{rajpurkar2016squad}, NewsQA~\cite{trischler2016newsqa}.\\
\textbf{Open Domain QA.} 
Open domain QA introduces large corpus as grounded documents and the task is to retrieve the appropriate documents and then try to extract the answer span. For this task, high-quality datasets have been proposed such as TriviaQA~\cite{joshi2017triviaqa}, WikiQA~\cite{cohen2018wikipassageqa}, QuAX~\cite{rashid2021quax}. Some previous work~\cite{kratzwald2018adaptive,lee2018ranking} selects answers from a closed set of passages or learns to rerank them. End-to-end open domain pipelines like DrQA~\cite{chen2017reading} and BERTserini~\cite{yang2019end} use TFIDF/BM25 for the retrieval of passages and a neural reader to select the answer span. ORQA~\cite{lee2019latent} and DPR~\cite{karpukhin2020dense} introduce a learnable retriever module with a dual encoder architecture.  
These works are all single-turn QA's whereas we target multi-turn conversations.\\
\textbf{Conversational QA.} Conversational QA is a variant of MRC where the queries are no longer single turn and the role of retrieval is disregarded~\cite{christmann2019look}. The multi turn questions can be interconnected (CoQA~\cite{reddy2019coqa}, DoQA~\cite{campos2020doqa}), can depend on the previous history answer(QuAC~\cite{choi2018quac}) or only limited to binary answers (ShARC~\cite{saeidi2018interpretation}). A better understanding of the context of conversation history is needed to answer the grounded question. To capture the context, FlowQA~\cite{huang2018flowqa} and GraphFlow~\cite{chen2019graphflow} use each word as nodes in a graph and use an attention mechanism to represent the history; HAE~\cite{qu2019bert} considers the history ground answers as context which is impractical for real life dialogue agents; Pos-HAE~\cite{qu2019attentive} considers the history turn positions as additional encoding. There are also backtracking based~\cite{qiu2021reinforced} and query rewriting based~\cite{vakulenko2021question, anantha2020open} models as mentioned in previous sections.\\
\textbf{Open Domain Conversational QA.} ODQA models like OrConvQA~\cite{qu2020open}, WS-OrConvQA~\cite{qu2021weakly}, ConvADR-QA~\cite{fang2022open} do not perform any history modeling and use the same history window in all of their modules for the end-to-end system. Other ODQA works like TopiOCQA~\cite{adlakha2022topiocqa} do not focus on history modeling and use gold training data to train neural models for their pipeline. Similar to \textit{Extractive QA} like this task (OrConvQA), there are \textit{Abstractive} pipelines~\cite{lewis2020retrieval} where the answer is generated using transformers like BART~\cite{lewis2019bart} rather than being extracted from passages. Such pipelines are sequence-to-sequence tasks and not span predictions.
\section{Conclusion}
\label{sec:conclusion}
We have presented the first end-to-end pipeline that uses non-uniform history modeling for open retrieval conversational question answering. We show that existing systems are suboptimal due to modeling the context in the same way for all modules and not utilizing previously returned passages for the \textit{Retriever} module. We have also proposed a novel algorithm to utilize such passages to output higher-quality passages for subsequent modules. We further updated the doc2dial dataset to make it appropriate for OrConvQA task. Extensive experimental evaluation from various history modeling techniques with different types of data shows that NORMY significantly outperforms the state-of-the-art in each individual module and the entire pipeline.

\section*{Acknowledgment}
This work was partially supported by NSF Grants IIS-1901379 and IIS-2227669.


\bibliographystyle{IEEEtran}
\bibliography{icsc}

\end{document}